\input{aipcheck}

\documentclass[
    ,final            
  ]
  {aipproc}

\layoutstyle{6x9}

\begin{document}

\title{Hydrodynamics of Internal Shocks in Relativistic Outflows}

\author{M. Kino}{
  address={SISSA, via Beirut 2-4, 34014 Trieste, Italy}
}

\author{A. Mizuta}{
  address={Yukawa Institute for Theoretical Physics, 
Kyoto University, Kyoto 606-8502, Japan}
}

\author{A. Celotti}{
  address={SISSA, via Beirut 2-4, 34014 Trieste, Italy}
}

\author{S. Yamada}{
  address={Science and Engineering, Waseda University, Shinjyuku, 
  Tokyo 169-8555, Japan}
}

\begin{abstract}
We study the hydrodynamical effects of two colliding shells, 
adopted to model internal shocks 
in various relativistic outflows such as gamma-ray bursts and blazars. 
We find that the density profiles are significantly affected by 
the propagation of rarefaction waves. 
A split-feature appears at the contact discontinuity of the two shells. 
The shell spreading with a few ten percent of the speed of light is also 
shown to be a notable aspect.
The conversion efficiency of the bulk kinetic energy to internal 
one shows deviations
from the widely-used inelastic two-point-mass-collision model. 
Observational
implications are also shortly discussed.
\end{abstract}

\maketitle

\section{Introduction}

The internal shock scenario proposed by Rees (1978) is 
one of the most promising models to 
explain the observational features of relativistic outflows
associated to e.g., gamma-ray bursts, and blazars 
(e.g., Rees \& Meszaros 1992; Spada et al. 2001).
Most of the previous works focus on the comparison 
of the model predictions 
employing a simple inelastic 
collision of two point masses 
with observed light curves 
(e.g., Kobayashi et al. 1997; 
Daigne \& Mochkovitch 1998) 
and little attention has been paid  
to hydrodynamical processes in the shell collision. 
However, it is obvious that, 
in the case of relativistic shocks,
the time scales in which 
shock and rarefaction waves cross the shells
are comparable to 
the dynamical time scale $\Delta^{'}/c$,
where $\Delta^{'}$ is the shell width measured in 
the comoving frame of the shell
and $c$ is the speed of light.
Since the
time scales of observations of these relativistic outflows 
(e.g., Takahashi et al. 2000 for blazar jet;
Fishman \& Meegan 1995 for GRBs)
are much longer than the dynamical one,
the light curves should contain the footprints of 
these hydrodynamical wave propagations.
Thus, it is worth to study the difference between 
the simple two-point-mass-collision 
(hereafter two-mass-collision) model
and the hydrodynamical treatment.
Here we show
(1) the hydrodynamical effects, 
especially
including the propagations of rarefaction 
wave (Kino, Mizuta \& Yamada 2004 hereafter KMY)
and
(2) some implications of hydrodynamical model
on the observed phenomena.

\section{Internal Shock Model}

Here we consider the
hydrodynamics of the two-shell interaction
(the fundamental physics of relativistic shocks  
can be found in Landau \& Lifshitz 1959 and Blandford \& McKee 1976).
Our main assumptions are 
(1) adopt a planar 1D shock and
neglect radiative cooling for simplicity,
(2) neglect the effect of magnetic fields,
and
(3) limit our attention to  shells with relativistic speeds.

\begin{figure}
  \includegraphics[height=.2\textheight]{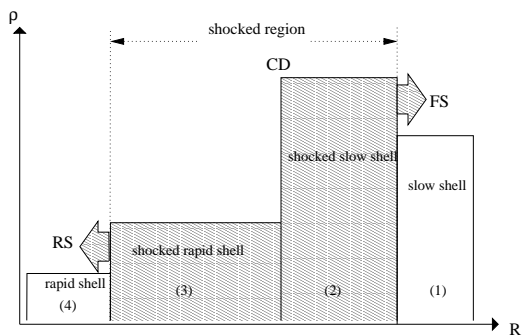}
  \caption{
Sketch of a two-shell-collision where
a rapid shell catches up with a slower one in 
the contact discontinuity (CD) rest frame (hereafter CD frame).
Forward and reverse shocks (FS and RS) propagate from the CD. 
We adopt the conventional numbering for each region used in the study
of GRB (e.g., Piran 1999).}
\label{fig:shock_is}
\end{figure}

In Fig. \ref{fig:shock_is}, 
we draw a schematic 
mass density profile of the shock propagation
during the interaction of a rapid (fast) and a slow shell.
Two shocks are formed: 
a reverse shock (RS) that moves 
into the rapid shell
and a forward shock (FS) that propagates into the slow one. 
There are four characteristic regions: 
(1) the unshocked slow shell,
(2) the shocked slow shell,
(3) the shocked rapid shell, and
(4) the unshocked rapid shell.
Thermodynamic quantities, 
such as 
rest mass density $\rho$, 
pressure $P$, 
and 
internal energy density $e$
are measured in the fluid rest frames.
We use the terminology 
of  {\it regions} $i$ ($i$=1, 2, 3, and 4) and
{\it position of discontinuity}
$j$ ($j$=FS, CD, and RS) where 
CD stands for a contact discontinuity.
The fluid velocity  
and Lorentz factor in the region $i$ 
measured in the ISM rest frame  
are expressed as
$\beta_{i}c$, and $\Gamma_{i}$, respectively.
%
%
Throughout this work, we use the assumption of $\Gamma_{i}\gg 1$.

We perform special 
relativistic hydrodynamical simulations.
The details of the code are given in
Mizuta et al. (2004)
and references therein.
We assume plane symmetry and consider the 
one dimensional motions of shells.
In discussing the propagation of shock and rarefaction waves, 
we choose the CD frame.
We start the calculation at $t=0$ when the collision of
two shells has just begun.
In this work,
we set $\Delta_{s}^{'}/c=1$ 
and  $\rho_{r}=1$ 
as units of the  numerical simulations. 
The subscripts $r$ and $s$ represent rapid and slow shells, respectively.
As for an equation of state (EOS), 
we set an adiavbatic index 
$\hat{\gamma}_{i}=5/3$ for non-relativistic case and 
$\hat{\gamma}_{i}=4/3$ for relativistic case, respectively. 
Initially, two shells have opposite velocities,
namely, $\beta_{\rm r}^{'} > 0$ and $\beta_{\rm s}^{'} < 0$.
We assume that the boundary of
each shell is kept intact
during the passage of 
shocks and rarefaction waves
(i.e. we do not impose any condition for the plasma
surrounding the two shells).


Bearing in mind the application to relativistic outflows
in GRBs and blazars,
we assume that the widths of two shells are
the same in the ISM frame, i.e.
 $\Delta_{\rm r}/\Delta_{\rm s}=1$ 
(see, e.g., Nakar \& Piran 2002 hereafter NP02).
It seems natural to suppose 
that ejected shells 
from the central engine have a correlation among them.
Here we consider following three cases; 
(1) the energy of bulk motion of the rapid shell ($E=\Gamma mc^{2}$) 
equals to that of the slow one in the ISM frame
(we refer to it as ``equal energy (or $E$)''),
(2) the mass of rapid shell ($m=\rho \Gamma\Delta$)
equals to that of the slow one 
(hereafter we call it ``equal mass (or $m$)''),
and 
(3) the rest mass density of rapid shell equals to that of the slow one 
(hereafter we call it ``equal rest mass density (or $\rho$)'').
Due to space limitations,
we highlight, in this proceeding,
the case of equal mass 
which satisfies
$\rho_{\rm r}\Delta_{\rm r}\Gamma_{\rm r}
=\rho_{\rm s}\Delta_{\rm s}\Gamma_{\rm s}$.

\section {Results}

\subsection{Shell ``split'' and ``spread''}

It is interesting to notice that,
when
%
%
the rarefaction wave 
from the FS side reaches the CD earlier
than the one from the RS side, then
the ``split'' occurs at the CD
since
the rarefaction wave going
from the higher density region (region 2)
into the
lower density region (region 3) 
makes a dip in the latter region.
The ``split'' feature is clearly
seen in Figs. 2 and 3.


In principle,
we can obtain
the speed of the rarefaction wave using Riemann invariants.
In the relativistic limit,
it is known that the speed of the head  
of the rarefaction wave is close to the speed of light (e.g., Anile 1989).
As the EOS of the shocked region deviates 
from the relativistic one, the speed is reduced.
The head propagation 
is identified as the shell ``spread''
which appears in Figs. 2 and 3.

\subsection{Energy conversion efficiency}

The conversion efficiency
of the bulk kinetic energy into internal one
is a fundamental issue.
The estimation
of the energy conversion
efficiencies with shock and rarefaction waves
taken into account are presented in 
Fig. \ref{fig:effmass},
on the basis of our 1D numerical simulations.
By analogy with the two-mass-collision model,
we define the efficiency measured in the ISM frame as
\begin{eqnarray}\label{eq:eff}
\epsilon(t)&\equiv&1-
\frac{ \int 
\Gamma(t,x)~dm(t,x)}
{\Gamma_{\rm r}m_{\rm r}
+\Gamma_{\rm s}m_{\rm s}}
\end{eqnarray}
where
$dm(t,x)$,
$\rho(t,x)$,
$\Gamma(t,x)$, and
$\Gamma(t,x) dx$,
are
the rest mass element,
the rest mass density, and
the Lorentz factor, 
measured in the ISM frame, 
and 
the length of the line element in the CD frame,
respectively.
%
In Fig. \ref{fig:effmass}, we compare the numerical results 
with the predictions by the
two-mass-collision approximation.  
After the shock waves break-out from the shells,
the conversion efficiency
is reduced by several $10\%$ from 
the estimate of the two-mass-collision model 
after several dynamical times.

\begin{figure}
  \includegraphics[height=.2\textheight]{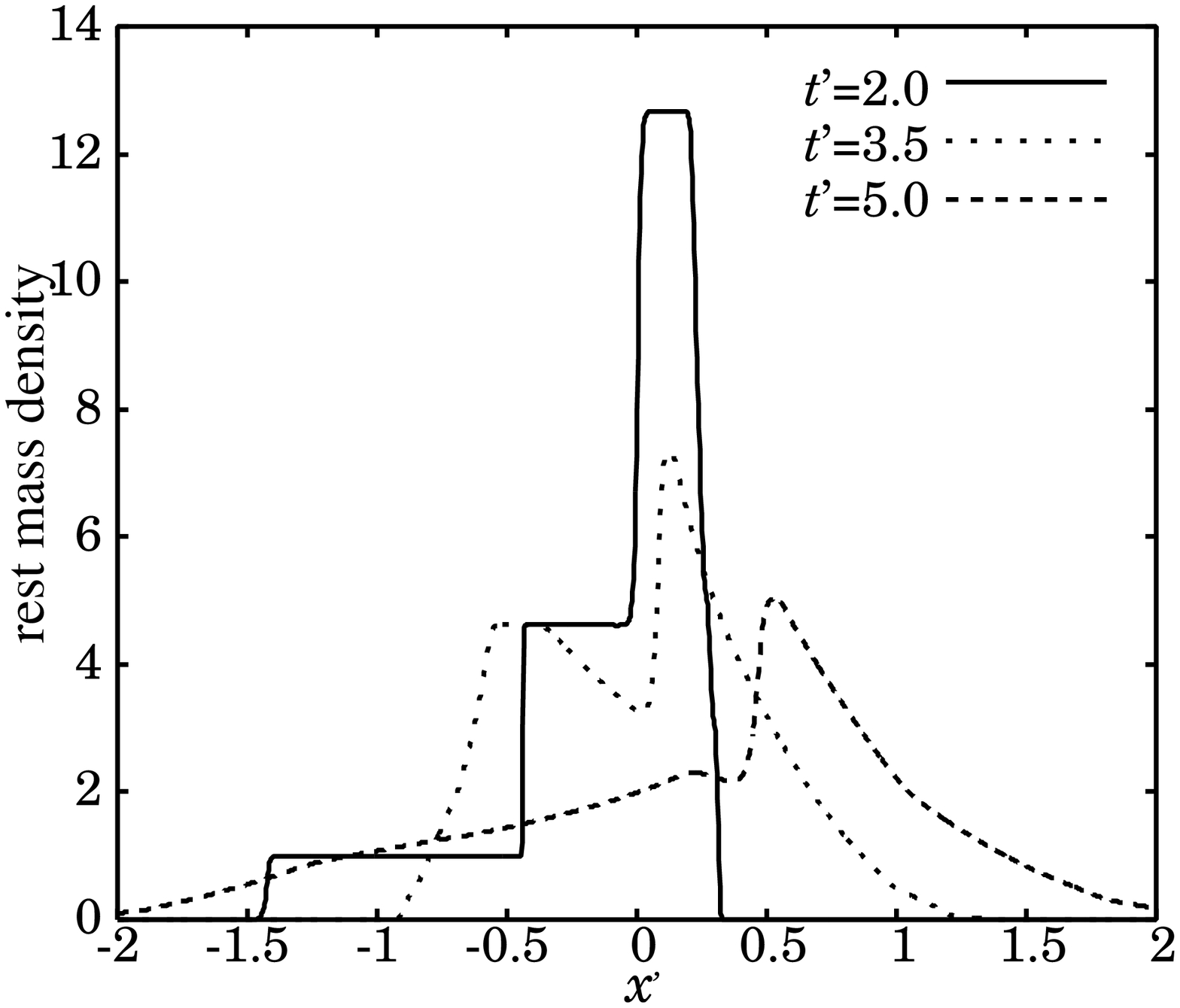}
\includegraphics[height=.2\textheight]{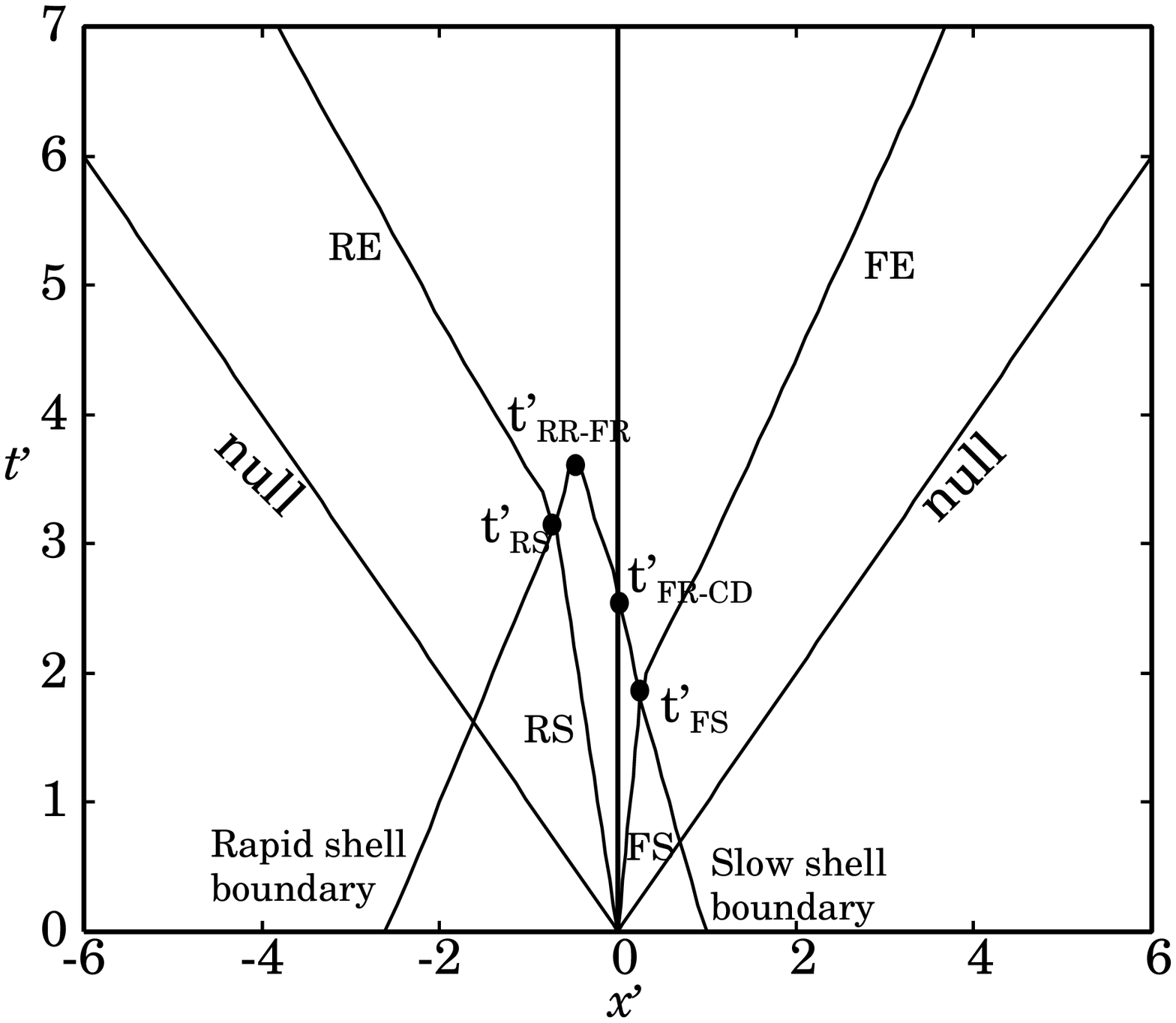}
  \caption{
Left:
Time evolution of 
the rest mass density profile in the CD frame for ``equal $m$''. 
In the ISM frame
$\Gamma_{\rm r}/\Gamma_{\rm s}=3$. 
The parameters are shown in Table 1.
Space-time diagram of shock and rarefaction waves propagations.
The rarefaction wave from the RS side (RE) 
spreads at the speed $\sim 0.8c$
while the one from FS side (FE) spreads at the speed $\sim 0.7c$.}
\end{figure}
\begin{figure}
  \includegraphics[height=.2\textheight]{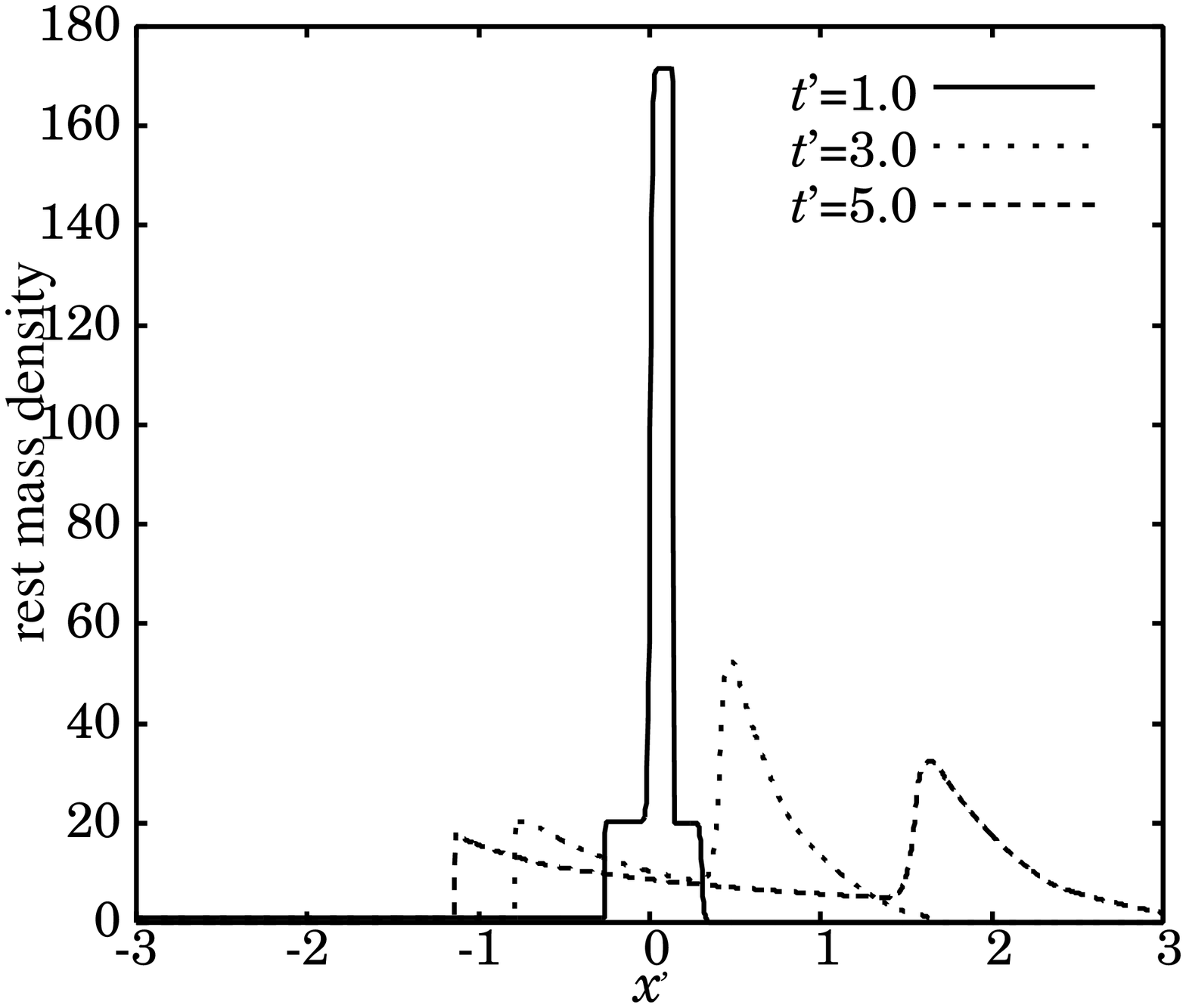}
\includegraphics[height=.2\textheight]{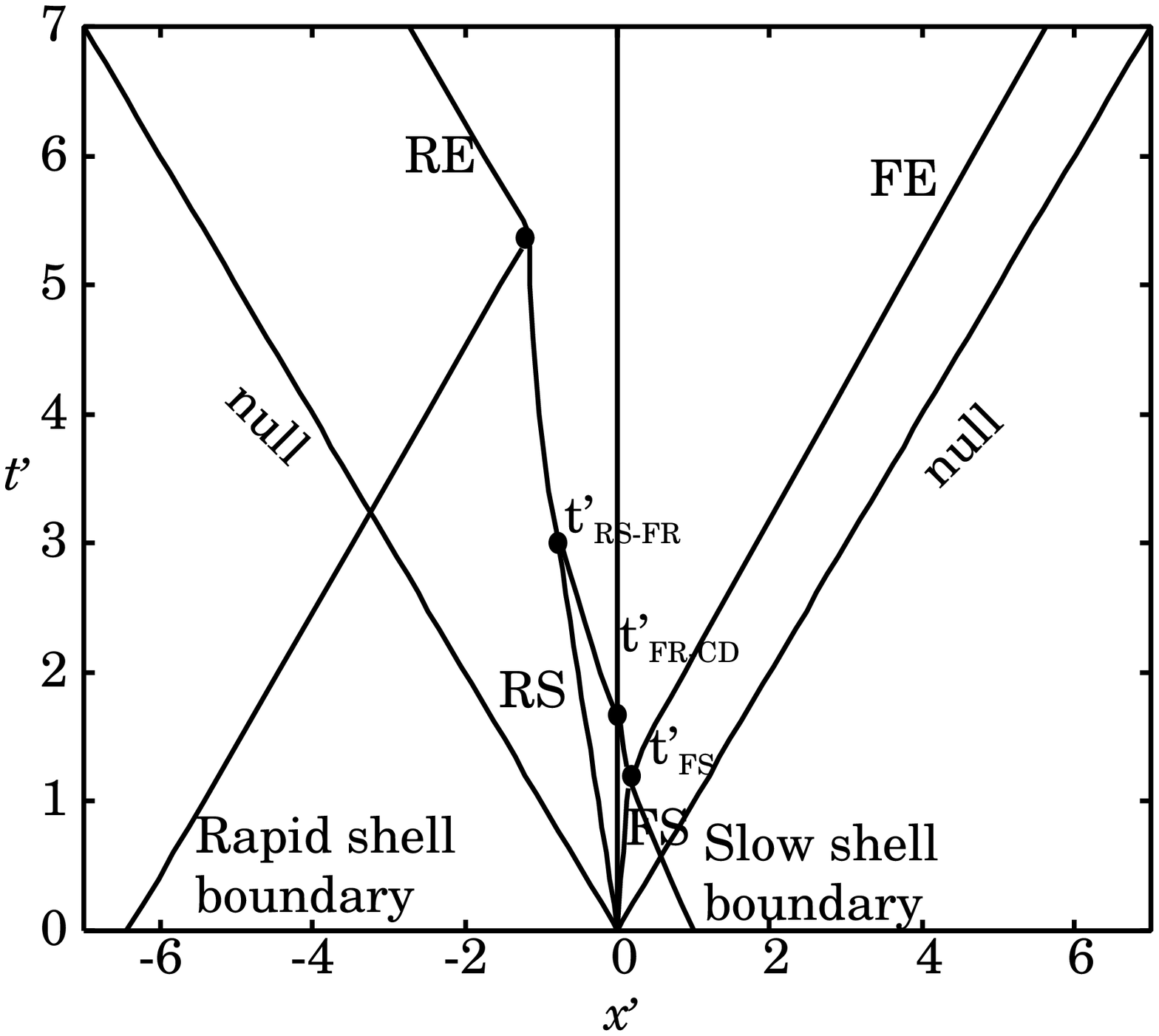}
  \caption{
Left:
Time evolution of the 
rest mass density profile in the CD frame for ``equal $m$''. 
In the ISM frame,
$\Gamma_{\rm r}/\Gamma_{\rm s}=20$. 
The parameters are shown in Table 1.
The low density rapid shell collides with the slow one 
and quickly spreads out. We also see that
the dense slow shock
is pushed forward by the rapid shell.
Right:
Space-time diagram of shock and rarefaction waves propagations.
RE spreads at the speed $\sim c$
while FE spreads at the speed $\sim 0.9c$.}
\end{figure}


\begin{table}
\begin{tabular}{lrrrrrr}
\hline
  & \tablehead{1}{r}{b}{$\Gamma_{\rm r}/\Gamma_{s}$ }
  & \tablehead{1}{r}{b}{$\Gamma_{\rm CD}$
\tablenote{
The slow shell Lorentz factor is fixed at $\Gamma_{\rm s}=5$. }}
  & \tablehead{1}{r}{b}{$\Gamma_{\rm r}^{'}$}   
  & \tablehead{1}{r}{b}{$\Gamma_{\rm s}^{'}$}   
  & \tablehead{1}{r}{b}{$\Delta_{\rm r}^{'}/\Delta_{\rm s}^{'}$}
  & \tablehead{1}{r}{b}{$\rho_{\rm s}/\rho_{\rm r}$}\\
\hline
equal $m$ (Fig.2)
& 3
& 7.6
& 1.25 
& 1.09
& 2.6
& 3
\\
equal $m$ (Fig.3) \tablenote{ We use 
$\hat{\gamma}_{3}=4/3$.} 
& 20
& 11.8
& 4.29
& 1.40
& 6.4
& 20
\\
\hline
\end{tabular}
\caption{
Parameter sets for numerical simulations 
of ``equal $m$''.}
\end{table}

\begin{figure}
  \includegraphics
[height=.25\textheight]{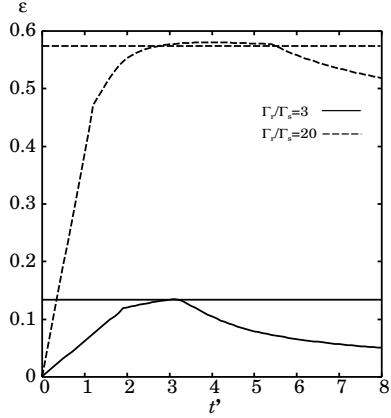}
  \caption{
Time evolutions of the  
conversion efficiency defined by Eq. (\ref{eq:eff})
for the ``equal $m$'' case.
During the shock propagations, 
$\epsilon$ 
approaches the constant values (horizontal lines)
estimated by two-point-mass model.}
\label{fig:effmass}
\end{figure}

\subsection{Observational implications}

It is worth discussing the hydrodynamic effects 
which may appears as observable features.
Here we discuss (i) the pulse shape in light curves, and (ii) the energy
conversion efficiency in multiple collisions.

\subsubsection{On  pulse shapes}

A pulse shape is basically determined by 
the rising and decaying times.
The relevant time scales that determine the pulse width are
(e.g., Piran 1999 and references therein): 
(i) The angular time, $ t_{\rm ang} $, which results
from the spherical geometry of the shells, 
(ii) The hydrodynamic time, $ t_{\rm dyn}
$, which arises from the shell's width and the shock crossing
time, and
(iii) The
cooling time - the time that it takes for the emitting electrons
to cool. 
If the cooling time 
is much shorter then $ t_{\rm ang} $ and $
t_{\rm dyn} $,
then no rarefaction wave is expected.
The hydrodynamics effect may however come from
the asymmetry of the crossing time of
forward and reverse shocks.   
When the cooling time 
is comparable  to $ t_{\rm ang} $ and $t_{\rm dyn} $,
then rarefaction waves may play a significant role on the 
pulse shape.
It is reasonable to suppose that the emission coming from the rarefaction
wave zone contributes to observed light curves. 

\subsubsection{On the conversion efficiency}

It is interesting to explore
the conversion efficiency
of the bulk kinetic energy into internal one
within the context of a realistic multiple collision case.
In Tanihata et al. (2002),
the conversion  efficiency of the internal shock scenario is assessed 
by evaluating the relative amplitude of flares 
as compared to the steady (``offset'') component.
Given the duration of the ejected shells,
the width of the pulse  
determines the hight of the ``offset'' component.
Wider width  make 
the ``offset'' shorter. 
As mentioned above,
the emission originating in the rarefaction
wave is likely to contribute to each observed pulse
in the weak cooling regime.


\section{Summary}

We have studied 
1D hydrodynamical simulations of two-shell-collisions  
in the CD frame. 
We find that rarefaction waves
have a dramatic effect on the dynamics.
Especially when a cooling time scale 
is sufficiently long in the shocked region, 
the observed emission
may significantly affected by them.

(1)
In the case of ``equal $m$'', the profile should 
in principle become
triple-peaked according to our classification scheme.
In practice, however, there is 
very short time for two-rarefaction-waves (FR-RR)
collision to produce a clear dip, 
while there is 
a lot more time for the FR
to create a dip over a fairly wide range of parameters. 
Therefore, the profile in this case
is effectively double-peaked (so-called shell ``split'').

(2)
For large $\Gamma_{\rm r}/\Gamma_{\rm s}$,
the ``spread'' velocity of the shells
after the collision is close to the speed of light.
Hence, the often used approximation
of constant shell width after collision
is not very good in treating multiple collisions (e.g.,  NP02).

(3)
The time-dependent 
energy conversion efficiency is quantitatively
estimated.
As the shell spreads after a collision, the internal energy 
is converted back into bulk kinetic energy 
due to thermal expansion.
If $\Gamma_{\rm r}/\Gamma_{\rm s}\gg1$
and the time-interval between collisions is long,
the conversion efficiency  
will substantially deviate from the estimate of
the two-mass-collision model.


\begin{theacknowledgments}

The work reported here was supported in part by
Grand-in-Aid Program for Scientific Research 
(14340066, 14740166, and 14079202) from the Ministry of Education,
Science, Sports, and Culture of Japan.
M. K. and A. C. acknowledge the Italian MIUR and INAF financial support. 

\end{theacknowledgments}


\bibliographystyle{aipproc}   

\bibliography{sample}

\IfFileExists{\jobname.bbl}{}
 {\typeout{}
  \typeout{******************************************}
  \typeout{** Please run "bibtex \jobname" to optain}
  \typeout{** the bibliography and then re-run LaTeX}
  \typeout{** twice to fix the references!}
  \typeout{******************************************}
  \typeout{}
 }

\end{document}